\documentclass[12pt,preprint]{revtex4-2}
\usepackage{amsmath}
\usepackage{hyperref}
\usepackage{graphicx}
\usepackage{float}
\usepackage{dcolumn}
\usepackage{bm}
\usepackage[english]{babel}
\usepackage{xcolor}

\begin{document}

\title{\textbf{Physics-Informed Neural Networks with Dynamical Boundary Constraints}
}

\author{Andrés Martínez-Esteban\textsuperscript{\hyperref[aff1]{1,}}\textsuperscript{\hyperref[aff2]{2}}}
\author{Pablo Calvo-Barl\'es\textsuperscript{\hyperref[aff2]{1,}}\textsuperscript{\hyperref[aff3]{3}}}
\author{Luis Mart\'in-Moreno\textsuperscript{\hyperref[aff2]{1,}}\textsuperscript{\hyperref[aff3]{3}}}
\author{Sergio G Rodrigo\textsuperscript{\hyperref[aff1]{1,}}\textsuperscript{\hyperref[aff2]{2}}}%
\email{Contact author: sergut@unizar.es}
\affiliation{%
\textsuperscript{1}\label{aff1}Departamento de Física Aplicada, Facultad de Ciencias, Universidad de Zaragoza, 50009 Zaragoza,Spain\\
\textsuperscript{2}\label{aff2}Instituto de Nanociencia y Materiales de Aragón (INMA),CSIC-Universidad de Zaragoza, 50009 Zaragoza, Spain\\
\textsuperscript{3}\label{aff3}Departamento de Física de la Materia Condensada, Facultad de Ciencias, Universidad de Zaragoza, 50009 Zaragoza,Spain
}%

\date{\today}

\begin{abstract}
 Physics-informed neural networks (PINNs) are numerical solvers that embed all the physical information of a system into the loss function of a neural network. In this way the learned solution accounts for data (if available), the governing  differential equations, or any other constraint known of the physical problem. However, they face serious issues, notably their tendency to converge on trivial or misleading solutions. The latter occurs when, although the loss function reaches low values the model makes incorrect predictions. These difficulties become especially significant in differential equations involving multi-scale behavior, such as rapidly varying terms and solutions exhibiting strong oscillatory behavior. To address these challenges, we introduce the Dynamical Boundary Constraint (DBC) algorithm, which imposes restrictions on the loss function based on prior training of the PINN. To demonstrate its applicability, we tested this approach on examples of different areas of physics.
 
\end{abstract}

\maketitle

\section{Introduction}
Differential equations are the backbone of calculus, as they describe the relationships between variables and their derivatives. They are essential to our comprehension of the natural world in physics areas and are extensively employed in other domains of knowledge. The challenge lies in the fact that most differential equations do not admit affordable analytical solutions, making it necessary to rely on numerical methods. 

With the widespread adoption of deep learning across various scientific fields, interest in its potential applications in solving differential equations has grown significantly, emerging as a promising alternative or as a complement to well-established numerical approaches. Early attempts to solve differential equations with Neural Networks (NN) date back to the 1990s~\cite{1998}. These methods relied on trial functions that satisfy boundary conditions, which had the drawback that each problem required different trial functions. More recently, new techniques have been introduced to address the difficulties of the pioneer methods, most notably the Physics-Informed Neural Networks (PINNs)~\cite{Raissi2019}. A PINN  is a type of NN that incorporates the underlying physical laws, typically expressed as differential equations, directly into the training process. Instead of learning from data alone, a PINN is trained to minimize a loss function that includes the residuals of the governing differential equations and related quantities (initial and/or boundary conditions), the error in observed data (if available) and, in general, any other constraint known of the solution. PINNs can incorporate the differential equations quite naturally, thanks to the capabilities of NNs to calculate derivatives using automatic differentiation.

PINNs are already applied across various fields, including fluid dynamics~\cite{Cai2021a}, quantum mechanics~\cite{cuantica3}, photonics~\cite{fotonica2,Chen2020}, or engineering studies~\cite{ZhangShaotong2024,Varey2024}. This underscores their applicability to a wide range of scientific fields, motivating numerous studies focused on their advancement and optimization ~\cite{whereweare,Farea2024}. 


However, PINNs still face significant challenges~\cite{Rathore2024}. One major issue is their tendency to converge to trivial (null) solutions when applied to differential equations over large domains. Additionally, they may yield misleading results that seem correct due to low total loss values achieved during the training process but are, in fact, incorrect. Such difficulties are especially evident in differential equations whose solutions or defining terms 
exhibit multi-scale values and/or strongly oscillatory behavior. 

Several approaches have been proposed to address the challenges faced by PINNs when dealing with differential equations that involve multi-scale terms or solutions exhibiting strong short-scale variations. These approaches include regularizing loss terms to maintain similar orders of magnitude between the different parts of loss functions or utilizing two NNs to separately solve the low-order and high-order oscillating terms~\cite {Wang2024}. Unbalanced gradients during backpropagation have been identified as a potential cause of PINN malfunction~\cite {Wang2021}, suggesting that adjusting the learning rate for individual loss terms may improve predictive accuracy.

Chaotic systems provide a typical example of differential equations with highly oscillatory solutions. PINNs have been shown to struggle with such systems, as reported in Ref.\cite{cheat}, where a PINN was applied to solve the double pendulum equation. Several strategies have been proposed to overcome this limitation. One notable approach involves including in the PINN architecture Fourier layers, which apply sinusoidal filters to the input. This modification improves the PINN capacity to represent and adapt to the rapid variations characteristic of highly oscillatory solutions~\cite{Wang2021a}. Another technique makes use of transfer learning, where the PINN is first trained on low-frequency problems and subsequently fine-tuned on high-frequency ones~\cite{Mustajab2024}. This approach has been shown to yield improved predictive accuracy for oscillatory equations, such as the damped cosine.

Finally, PINNs tend to independently satisfy initial conditions and reduce the loss that accounts for the differential equation without guaranteeing consistent solutions across the entire domain~\cite{NEURIPS2021_df438e52}. One possible solution involves dividing the domain into smaller subintervals, with the PINN trained separately on each subinterval. However, this methodology considerably constrains the capacity of the PINN to generalize across the entire domain.

Here, we present the Dynamic Boundary Constraint (DBC) algorithm as a solution to the above-described limitations of PINNs. The core principle of the algorithm lies in the incorporation of additional restrictions, interpreted as boundary conditions, which are dynamically integrated into the training process alongside the existing initial condition, thus enabling the PINN to ensure accuracy throughout the domain.

This paper is structured as follows. Section \ref{DBC-section} illustrates how the standard PINN framework struggles to accurately solve even a basic ordinary differential equation (ODE), such as the harmonic oscillator. In contrast, the proposed DBC method yields a solution that closely matches the correct result. In Section~\ref{ray-section}, we illustrate how DBCs effectively work even though the problem includes multi-scale terms. As an illustration, we apply the DBC method to model the trajectory of light near a massive object using the optical ray equation. In Section~\ref{lorenz-section}, we illustrate the use of the DBC method on the Lorenz equations, which show significant oscillatory behavior across large domains. Finally, we conclude with a summary of our findings.

All PINNs to conduct this work have been implemented using the Keras and TensorFlow libraries~\cite{Chollet2022}. The code related to this paper can be found elsewhere~\cite{codigo}.

\section{Dynamic Boundary Constrains}\label{DBC-section}
We expose the problems that PINNs face when applied to large domains by examining the one-dimensional harmonic oscillator: 
\begin{equation}
\frac{d^2y}{dx^2}=-y,
\label{harmonic oscillator eq}
\end{equation}

The PINN will receive as input the different values of $x$ and will output its prediction for the value $y(x)$. The PINN loss function includes two terms: one for the differential equation ($L_D$) and another for the initial conditions ($L_{IC}$):

\begin{equation}
L_D=\frac{1}{N}\sum_{i=1}^N\left(\frac{dy_{NN}^2}{dx^2}(x_i)+y_{NN}(x_i) \right)^2
\label{LD harmonic}
\end{equation}

\begin{equation}
L_{IC}=(y_{NN}(x_0)-y(x_0))^2+(\frac{dy_{NN}}{dx}(x_0)-\frac{dy}{dx}(x_0))^2
\label{LIC harmonic}
\end{equation}

The subscript NN indicates the predictions made by the PINN. Eq.\ref{LD harmonic} evaluates these parameters at a set of $N$ distinct points $x_i$, which corresponds to the training data.  In Eq.\ref{LIC harmonic}, the output of the PINN is evaluated at the coordinate corresponding to the initial condition, $x_0$. 

The chosen PINN architecture has three hidden dense layers, each containing 500 neurons. The trainable parameters are initialized using the Glorot Uniform initializer, and the activation function employed is the hyperbolic tangent in all cases. We use the Adam optimizer with a learning rate of $5 \times 10^{-5}$ for optimization. This setup will be used throughout this work unless otherwise stated.

\begin{figure}[h]
\centering
\includegraphics[width=1\linewidth]{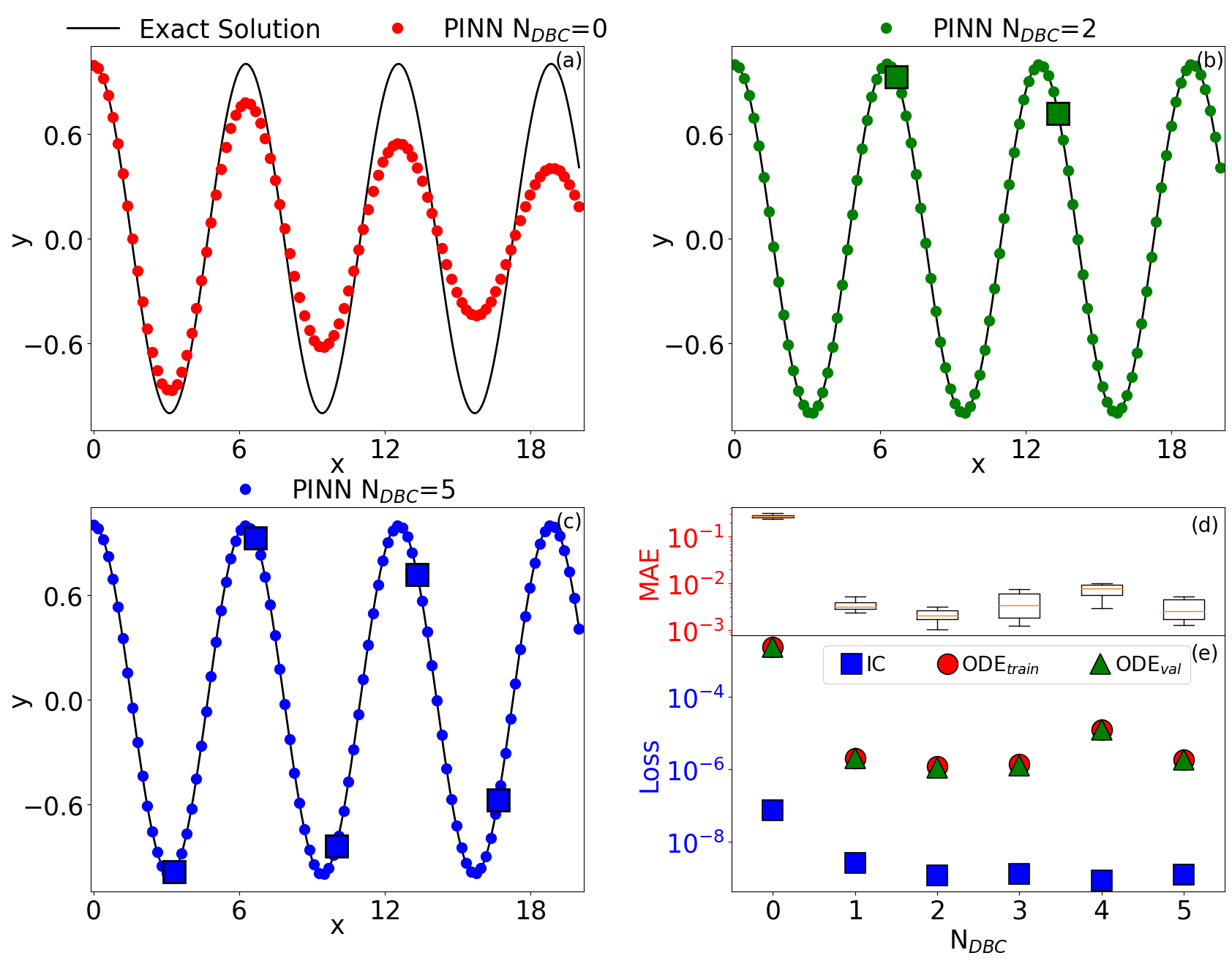}
\caption{\scriptsize Comparison of PINN predictions (dots) with the exact solution (solid line) for the harmonic oscillator equation: (a) standard PINN ($N_{DBC}=0$); (b) PINN with two DBC (square marks their location); and (c) PINN with five DBCs. (d) Box-plot of the mean absolute error (MAE) between the PINN predictions and the exact solution over ten independent training runs, as a function of $N_{DBC}$. The orange line denotes the median, while the whiskers represent the minimum and maximum errors. The PINN results shown in panels (a)–(c) correspond to the lowest observed error. (e) Training losses corresponding to the lowest-error model for each DBC configuration in (d). The initial condition loss, $L_{IC}$, is shown with squares. The ODE loss, $L_{D}$, is evaluated at $N_{\text{train}} = 200$ training points (circles) and at validation points (triangles), with the latter chosen such that $N_{\text{val}} = 10 \times N_{\text{train}}$. 
}
\label{harmonic oscillator multiplot}
\end{figure}

Figure~\ref{harmonic oscillator multiplot}(a) demonstrates that a standard PINN without DBCs (red dots) does not successfully predict the solution to the harmonic oscillator equation $y_{exact}= \cos(x)$ (black line) across a domain covering multiple periods. The analytical solution is obtained for the initial conditions $y(x_0)=1$ and $y'(x_0)=0$ at $ x_0=0$. We take $N=200$ points for training, which are regularly distributed within the interval $x\in[0,20]$. Though trained on this interval, the PINN has difficulties capturing the harmonic behavior, leading to significant deviations from the actual solution, particularly as $x$ increases. This highlights the limitations of the standard PINN when applied to ODEs with solutions that exhibit oscillatory responses over broad domains.

An interesting tendency is observed in the prediction made by the standard PINN. Near the region where the initial condition is imposed, there is excellent agreement between the analytical solution and the prediction of the PINN. However, as we move further from this point, the trend begins to deviate. In this situation, the solution asymptotically approaches the zero function, which, while valid, is not the solution to Eq.~\ref{harmonic oscillator eq} for the specified initial conditions.

The core concept of the DBC method is to enhance the generalization capabilities of the PINN by incorporating additional constraints into the loss function. The procedure is as follows: the interval over which the ODE is to be solved by the PINN is partitioned into subintervals. Training begins on the subinterval containing the coordinate of the initial condition, $x_0$. Once training on the current subinterval is complete, the prediction of its endpoint is saved and incorporated as an additional constraint in the loss function. Training then proceeds to the next subinterval. At this stage, two approaches may be employed: (i) a cumulative approach, which includes training data from both the current and all previous subintervals, or (ii) an individual subinterval approach, where the PINN is trained exclusively on data within the new subinterval. Regardless of the chosen approach, the final training phase is performed over the entire domain. 

Additionally, our tests suggest that PINNs tend to perform better when their trainable parameters are randomly reinitialized before training on each new subinterval. This improvement is likely related to the PINN tendency to become trapped in local minima, which can hinder its adaptation to newly introduced subintervals. However, this effect is not a decisive factor, as PINNs can still achieve satisfactory performance without reinitialization.

The loss term representing all the constraints to the loss different from the differential equation, that is, the initial condition (IC) and the DBCs, is given by:
\begin{align}
&L_{C}=L_{IC}+\sum_{\alpha=1}^{N_{DBC}}(y_{NN}(\hat X_{\alpha})-\hat Y_\alpha)^2,
\label{new LB}
\end{align}
where $\hat Y_\alpha$ is the saved PINN prediction at the endpoint $\hat X_\alpha$ of the $\alpha$‑th training interval, which we refer to as a DBC, and $N_{DBC}$ is the number of DBCs. 
 
The impact of introducing DBCs is demonstrated in Fig.~\ref{harmonic oscillator multiplot} (b) and (c), with two and five DBCs (represented by squares), the PINN predictions (shown as dots) perfectly match the exact solution. These results underscore the effectiveness of incorporating DBCs in constraining the standard PINN model, enabling it to get the correct solution within the domain where the ODE is intended to be calculated. Note that the cumulative approach was used in this example, but the individual subinterval approach can also yield accurate solutions, as demonstrated in Fig.~1 of the Supplementary Material. 

At this point, the deviation between the exact solution and the PINN prediction becomes evident in the standard PINN case, highlighting a critical limitation that is effectively addressed by incorporating DBCs. This improvement is statistically supported, as shown in the box plots of Fig.~\ref{harmonic oscillator multiplot}(d) from the Mean Absolute Error (MAE) distribution of ten independent trainings for each N$_{DBC}$. The error bars (whiskers) indicate the maximum and minimum values of the error. The orange line indicates the median of the distribution.

At first glance, the error observed in standard PINNs might appear to stem from difficulties in minimizing the loss during training. However, this is not the case. As shown in Fig.~\ref{harmonic oscillator multiplot}(e), the losses corresponding to the initial conditions $L_{IC}$ (squares) are very low in the standard PINN configuration ($L_{IC} \sim 10^{-7}$). The losses associated with the differential equation, $L_D$, remain low ($\sim 10^{-3}$) regardless of whether they are computed on the training or validation dataset (a denser dataset with ten points between each training pair). The absence of overfitting in this case is further confirmed by Fig.~2 (a) ($N_{DBC}=0$) and (b) ($N_{DBC}=5$) of the Supplementary Material, which plot $L_D$ as a function of $x$ for both datasets. This relatively small value would indicate a good performance of the PINN, considering that the oscillatory terms in Eq.~\ref{harmonic oscillator eq} are generally of larger magnitude. This showcases that low-loss values do not necessarily guarantee accurate PINN predictions. This can be attributed to a common behavior in PINNs: once the initial conditions are satisfied, $L_D$ decreases across all points, even if the resulting local solutions do not truly correspond to the given initial conditions.

Note that the number of DBCs, despite the intuitive assumption that more is better, cannot be reliably used as an indicator of the accuracy of the solution. As depicted in Fig.~\ref{harmonic oscillator multiplot}(d), the associated error (minimum value in the box plot) increases when moving from $N_{DBC}=3$ to $N_{DBC}=4$, even though, as shown in panel (e), this change slightly reduces $L_{D}$. This trend is even more obvious in later examples.

To further explore the potential of the DBC method, we will apply it to more challenging problems. First, to illustrate a multiscale problem, we will model the trajectory of a light ray in curved spacetime near a massive gravitational object (e.g., a black hole) using the optical ray equation. Second, we will investigate the Lorenz equations, the most iconic dynamical system that laid the foundation for chaos theory. These examples serve to evaluate the robustness of our method on ODEs exhibiting multi-scale and high oscillatory behavior while also showing its applicability to diverse areas of physics.

\section{Case studies and applications}

\subsection{The optical ray equation} \label{ray-section}
The Fermat principle remains fundamental in geometric optics despite being formulated centuries ago. It inspires innovative applications, such as the recent development of generalized laws of reflection and refraction in metasurfaces~\cite{Yu2011}. Another example of the method's relevance is the study of light trajectories in GRaded-INdex (GRIN) media using the optical ray equation derived from the principle of minimal action. GRIN lenses offer significant advantages over traditional lenses, including their ability to miniaturize optical systems through compact designs while maintaining high-resolution imaging and efficient light transmission. The optical ray equation is also valuable for designing mini-endoscopes in the study of deep zones of the brain~\cite{Bagramyan2019,Tabourin2024}, creating lightweight and miniaturized lenses~\cite{Xiao2023}, and performing inverse design of the refractive index distribution from a predetermined light path~\cite{Zhang2024}. The optical ray equation is analytically affordable in only a few simple cases. Heterogeneous distributions of the refractive index and the realistic optical properties of materials (e.g., dispersion) often lead to differential equations for which analytical solutions are either unknown or nonexistent. In such cases, numerical methods or approximations are essential to predict the solutions~\cite{Moore1975,Sharma1982,Ohno2020}.


We applied the standard PINN enhanced with the DBC method to the optical ray equation. Exemplarily, we calculated light paths around massive stellar objects (e.g., black holes), where the gravitational pull on light is modeled as an effective refractive index~\cite{Chatak2004}, based on the Schwarzschild metric~\cite{Nolte2019},
\begin{equation}
n(r)=\frac{1}{1-\frac{r_s}{r}}
\label{refractive index}
\end{equation}
Here, $r_s=\frac{2GM}{c^2}$ is the Schwarzschild radius, which is composed of the gravitational constant $G$, the speed of light in vacuum $c$, the mass $M$, and the radial distance to the center of the astronomical object $r$. 

To describe the light path $\vec r$ in such an effective medium, we use the optical ray equation,
\begin{equation}
\frac{d}{ds}\left(n\frac{d\vec r}{ds}\right)=\nabla n
\label{ray equation}
\end{equation}
, where the parameter $s$ represents the arc-length along the ray path.

This is a clear example of multi-scale behavior in ODEs, as evidenced by the dependence of the path light on the gradient of $n$, which strongly varies near the Schwarzschild radius.

By limiting the problem to a plane that includes the massive object situated at the origin of the coordinate system and by expanding Eq.~\ref{ray equation}, we can express the ray equation for the trajectory $\vec r=(x,y)$  in Cartesian coordinates as follows:
\begin{eqnarray}
\frac{d p_x}{d s}=\frac{\partial n(x,y)}{\partial x}=-n^2\frac{r_sx}{r^3} \nonumber \\
\frac{d p_y}{d s}=\frac{\partial n(x,y)}{\partial y}=-n^2\frac{r_sy}{r^3},
\label{rayequation}
\end{eqnarray}
where $r=+\sqrt{x^2+y^2}$ and the momentum $\vec p=(p_x,p_y)=n\frac{d \vec r}{d s}$. Therefore:
\begin{eqnarray}
\frac{dx}{ds}=\frac{p_x}{n(x,y)} \nonumber \\
\frac{dy}{ds}=\frac{p_y}{n(x,y)}
\label{momentum}
\end{eqnarray}

To solve Eqs.~\ref{rayequation} and \ref{momentum}, we define a PINN whose input layer takes the arc-length parameter $s$, and whose output layer represents the four-dimensional vector $(x, y, p_x, p_y)$. The total loss function of the PINN is defined as $L = L_{D} + L_{C}$, where $L_D$ accounts for the contributions from Eqs.~\ref{rayequation} and \ref{momentum} (i.e., the differential equation residuals), while $L_{C}=L_{IC}+L_{DBC}$, the contributions from the initial conditions and DBCs respectively. Explicitly:
\begin{align}
&L_D =\frac{1}{N}\sum_{i=1}^N \bigg(a_1\cdot\left[r^3\frac{d{p_x^{NN}}}{d{s}}(s_i)+n^2r_sx^{NN}(s_i) \right]^2+
\nonumber \\
&a_2\cdot\left[r^3\frac{d{p_y^{NN}}}{d{s}}(s_i)+n^2r_sy^{NN}(s_i) \right]^2+
\nonumber \\
& a_3\cdot\left[\frac{d{x^{NN}}}{d{s}}(s_i)-\frac{p_x^{NN}(s_i)}{n}\right]^2+
\nonumber \\
&a_4\cdot\left[\frac{d{y^{NN}}}{d{s}}(s_i)-\frac{p_y^{NN}(s_i)}{n}\right]^2\bigg)
\label{LD}
\end{align}

\begin{align}
&L_{IC}=\nonumber \\
&b_1\cdot[x^{NN}(s_0)-x_0]^2+b_2\cdot[y^{NN}(s_0)-y_0]^2+
\nonumber \\
&b_3\cdot[p_x^{NN}(s_0)-p_{x_0}]^2+b_4\cdot[p_y^{NN}(s_0)-p_{y_0}]^2
\label{LIC}
\end{align}
The same mathematical expression as in Eq.~\ref{LIC} is used for $L_{DBC}$, where the phase-space vectors $\vec{\xi} = (\vec{r}, \vec{p})$ belong to the set of DBCs.

The super index NN states for the predictions (outputs) of the PINN and $n=n(x^{NN},y^{NN})$. Eqs.~\ref{rayequation} have been rearranged in Eq. \ref{LD} to avoid singularities for $r \rightarrow 0$. Equation~\ref{LIC} is evaluated at the initial value of the parameter $s$ ($s_0$), whereas Eq.~\ref{LD} is evaluated at $s_i \ (i = 1, \dots, N)$, corresponding to the training points. The hyperparameters $a_i$ and $b_i$ (i=1,...4) can be tuned to improve the performance of the PINN. In our case, $a_1 = 5$ and $a_2 = 10$, with all other values equal to one. This adjustment was based on initial tests, indicating that these components were more challenging to predict than other terms in the loss function. 

Regarding the NN, we used the same hyperparameters as those for the harmonic oscillator, except that here there are five dense hidden layers, each with 500 neurons. Similarly to the harmonic oscillator, the cumulative approach is employed. However, the individual subinterval approach is also capable of producing accurate solutions, as illustrated in Fig.~4 of the Supplementary Material.
 
Figure~\ref{Runge_kutta}(a) shows the light paths computed using the PINN (dots) and compares them to calculations with the fourth-order Runge-Kutta (RK) method (solid lines). We choose $r_s=1$ so that the light trajectory is neither perfectly straight nor it collapses into the massive object within the region of space under consideration. Two distinct impact parameters have been chosen (in units of $r_s$): i) $(x_0, y_0) = (-12.5, 6.25)$, and $(x_0, y_0) = (-12.5, 4.375)$. In both cases, the initial momentum is the same: $(p_{x_0}, p_{y_0}) = (12.5, 0)$. In this case, $N=200$ training points are used, evenly spaced within the range $s \in [0, 2]$. This range is large enough to capture the curvature of light induced by the massive object. The central circle represents the radius of the event horizon of the massive object. 

The PINN performs well with $N_{DBC} = 0$ at $y_0 = 6.3$ because this trajectory of light is only minimally affected by the presence of the massive object. However, the agreement deteriorates as the impact parameter reduces, such as for $y_0 = 4.4$. For this scenario, the RK method predicts more pronounced light bending near the massive object than the PINN prediction without DBCs. As the light path approaches the massive object horizon, gravitational forces rapidly intensify through the source terms in Eq.~\ref{rayequation}, creating an imbalance between the ODE terms, which ultimately leads to inaccurate predictions by the standard PINN.

Returning to Fig.~\ref{Runge_kutta}(a), we see the impact of applying the DBC method during training for the trajectory with $y_0 = 4.4$. The inclusion of DBCs, represented by triangles and squares, mitigates the discrepancies observed when the trajectory approaches the singularity.

To statistically validate the results, we trained the PINN ten times for each N$_{DBC}$ configuration corresponding to the ray path with the strongest bending ($y_0 = 4.4$). The results are presented in Fig.~\ref{Runge_kutta}(b), where a box plot summarizes the statistical distribution of the error (MAE) between the PINN predictions and the RK method, for each N$_{DBC}$. Overall, agreement with the RK results tends to improve as the number of DBCs increases, although not monotonically, a similar behavior as observed in the harmonic oscillator study. This is evident upon inspection of all the trajectories obtained with different DBCs (see Figs.~4 (a) and (b) of the Supplementary Material): again, no clear correlation is observed between the error and the loss values. Figure~5 of the Supplementary Material displays the error as a function of $L_{D}$ for the validation points, showing that the lowest error and best agreement between the PINN prediction and the RK method occur at $N_{DBC} = 7$.

In Fig.~\ref{Runge_kutta}(c), the losses $L_D$ and $L_{IC}$ corresponding to the lowest errors achieved during all training runs are indicated by triangles (validation data), circles (training data), and squares (initial conditions). The ODE loss remains consistently low ($< 10^{-4}$) across all cases, including the standard PINN, reinforcing the earlier observation from the harmonic oscillator study that a low loss value does not necessarily guarantee an accurate solution to the differential equation.

\begin{figure}[h!]
\centering
\includegraphics[width=0.7\linewidth]{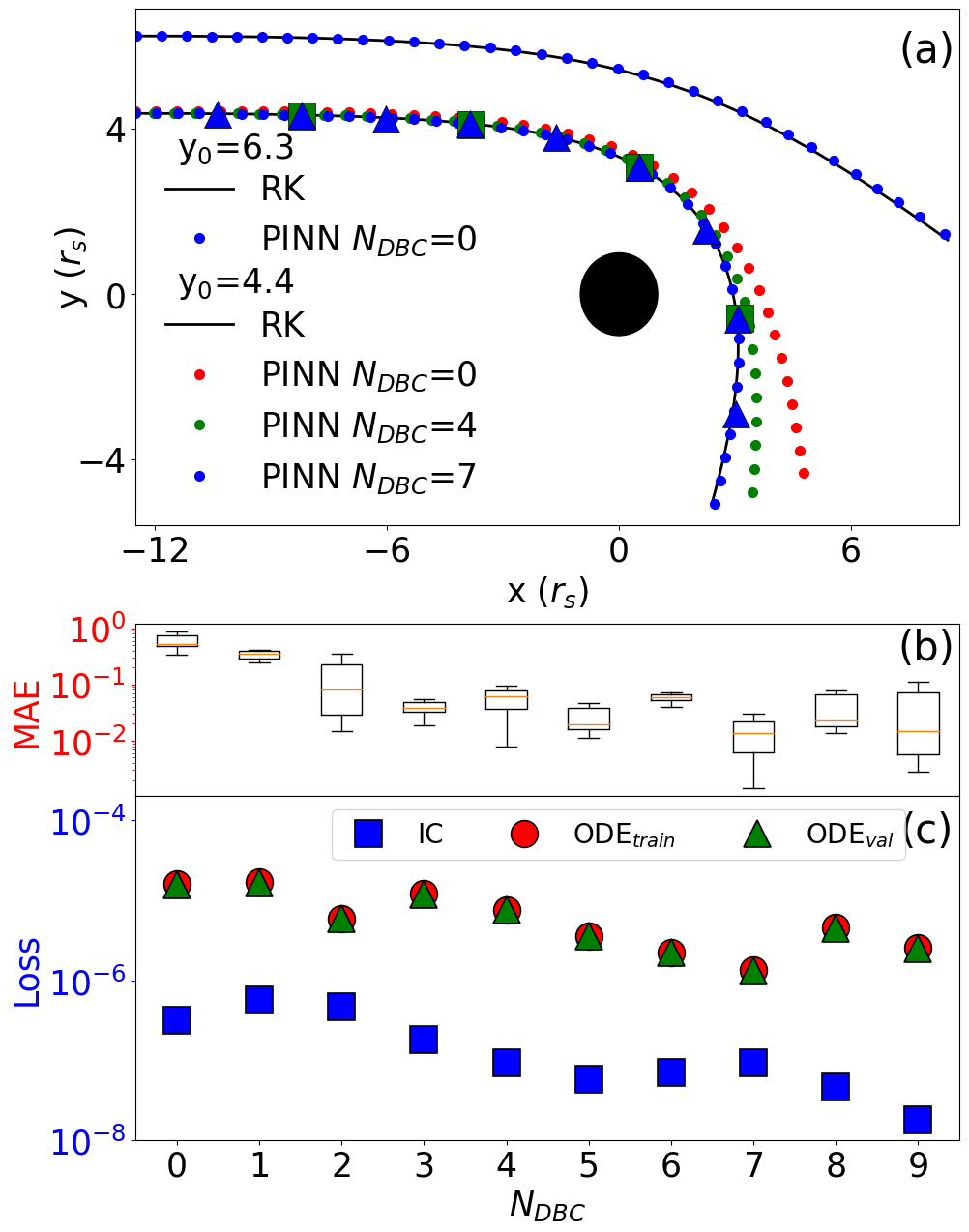}
\caption{\scriptsize (a) Comparison of light trajectories between the PINN predictions (dots) and the RK solutions (solid lines) for two initial conditions. Triangles and squares mark the locations of the DBCs for two different examples, while the central circle indicates the radius of the event horizon of the massive object. (b) Box-plot of the mean absolute error (MAE) between the PINN predictions and the RK method over ten training runs, as a function of $N_{DBC}$. The error bars (whiskers) represent the minimum and maximum error values. The orange line is the median of the distribution. (c) Training losses corresponding to the lowest-error model for each DBC configuration in (b). The initial condition loss, $L_{IC}$, is shown with squares. The ODE loss, $L_{D}$, is evaluated at $N_{\text{train}} = 200$ training points (circles) and at validation points (triangles), with the latter chosen such that $N_{\text{val}} = 10 \times N_{\text{train}}$}
\label{Runge_kutta}
\end{figure}

\subsection{Chaotic systems}
\label{lorenz-section}
The Lorenz equations, originally conceived as a model to study atmospheric convection, revealed a system of three-dimensional differential equations exhibiting chaotic dynamics ~\cite{Lorenz}. One of the most characteristic behaviors in chaotic systems is the appearance of regions in phase space toward which the system's trajectories converge, known as attractors, which present a strongly oscillatory dynamic. These equations are:
\begin{align}
\frac{dy_1}{dx} &= a \cdot (y_2 - y_1) \nonumber \\
\frac{dy_2}{dx} &= y_1 \cdot (b - y_3) - y_2 \nonumber\\
\frac{dy_3}{dx} &= y_1y_2 - cy_2,
\label{eq_lorenz}
\end{align}
where $a$ is the Prandtl number, and $b$ is the Rayleigh number.

The corresponding PINN consists of a single input neuron for $x$ and three output neurons representing the predictions of $y_1$, $y_2$, and $ y_3$.
The PINN's loss function is defined as before, $L = L_{D} + L_{C}$, where $L_D$ includes the contributions from Eq.~\ref{eq_lorenz}, and $L_{C}=L_{IC}+L_{DBC}$ accounts for the contributions from both the initial conditions and DBCs. Again, $L_{DBC}$ is equivalent to $L_{IC}$ but is applied at the locations of the DBCs.

\begin{align}
&L_D =\frac{1}{N}\sum_{i=1}^N \bigg(\alpha_1\left[\frac{dy_1^{NN}}{dx}(x_i)-a\cdot(y_2^{NN}(x_i)-y_1^{NN}(x_i))\right]^2+
\nonumber \\
&\alpha_2\left[\frac{dy_2^{NN}}{dx}(x_i)-y_1^{NN}(x_i)\cdot(b-y_3^{NN}(x_i)) +y_2^{NN}(x_i)\right]^2+
\nonumber \\
&\alpha_3 \left[\frac{dy_3^{NN}}{dx}(x_i)-y_1^{NN}(x_i)y_2^{NN}(x_i)+cy_2^{NN}(x_i) \right]^2\bigg)
\label{LD chaos}
\end{align}

\begin{align}
&L_{IC}(x_0)=
\beta_1[y_1^{NN}(x_0)-y_1(x_0)]^2+
\nonumber \\
&\beta_2[y_2^{NN}(x_0)-y_2(x_0)]^2+\beta_3[y_3^{NN}(x_0)-y_3(x_0)]^2
\label{LIC Chaos}
\end{align}

Consistent with previous notation, $x_i$ refers to the training coordinates, while $x_0$ denotes the initial point. The loss terms are weighted by $\alpha_i = 1$ and $\beta_i = 6000$ to prioritize the accurate enforcement of the initial conditions, which is critical for chaotic systems where small errors in the initial state can grow exponentially.

We used the same hyperparameters and NN architecture as for the harmonic oscillator, except for the addition of a Fourier layer before the dense layers, which improves convergence as in chaotic systems~\cite {Wang2021a}.

Given the complex behavior of the Lorenz dynamical system, we use $N_{DBC} = 800$ , since using a similar number to that in previous examples leads to prediction failure before the onset of the chaotic behavior. Note, however, that no attempt has been made to determine the optimal number of DBCs.

To reduce the computation time required for solving the equations with such a large number of DBCs, we use the subinterval DBC method described earlier, as it is faster than the cumulative approach.

\begin{figure}
\centering
\includegraphics[width=0.7\linewidth]{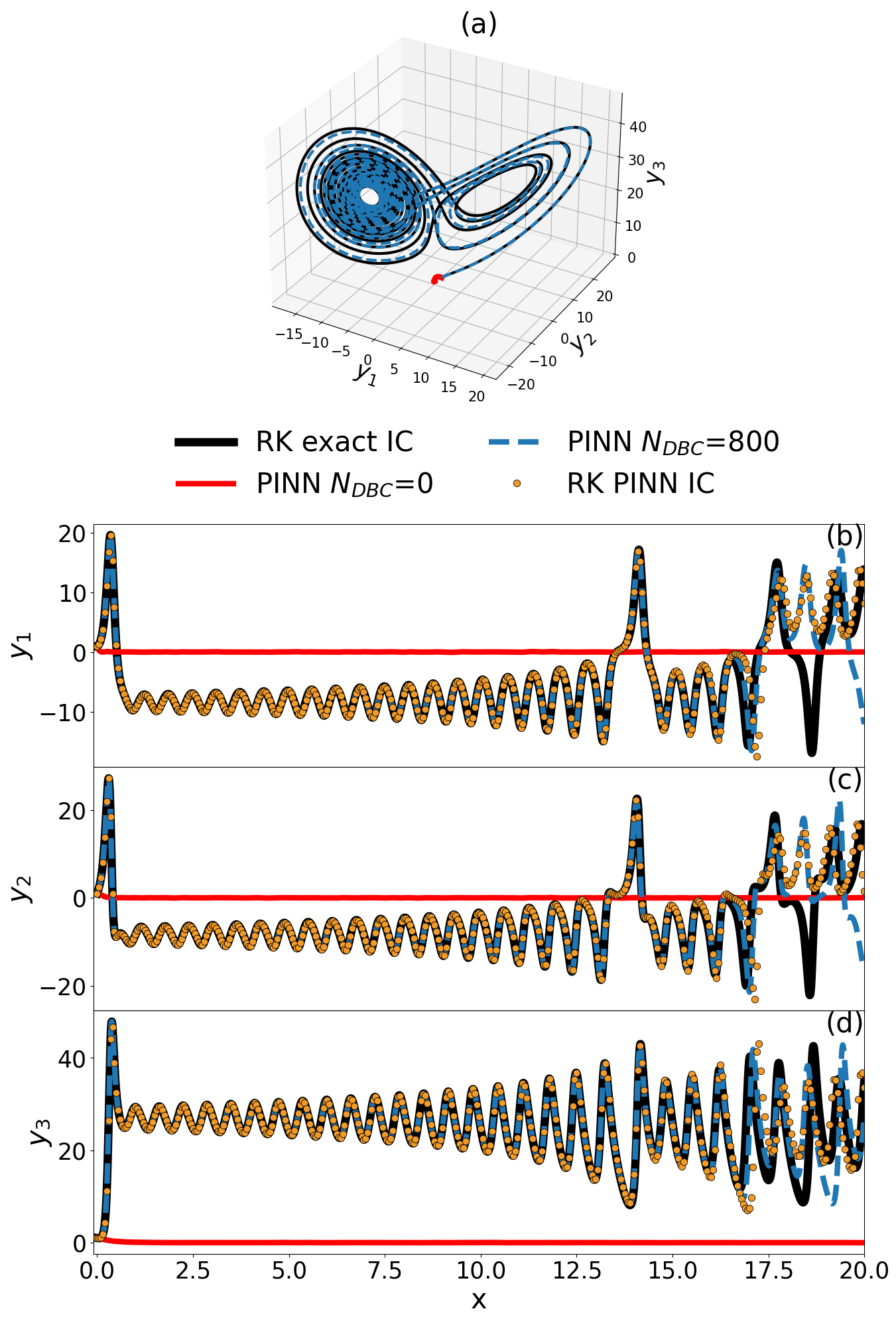}
\caption{\scriptsize   (a) Three-dimensional trajectory of the Lorenz attractor showing the RK solution (black line), the PINN prediction with DBCs (blue line), and the standard PINN prediction without DBCs (red line). (b)–(d) Individual components of the trajectory. The solid black line represents the RK solution with the exact initial conditions, the solid red line shows the standard PINN prediction, and the dashed blue line corresponds to the PINN with $N_{DBC} = 800$. The dotted orange curve depicts the RK solution obtained using the initial conditions predicted by the PINN.
}
\label{fig:runge_kutta_vs_pinn}
\end{figure}

Figure~\ref{fig:runge_kutta_vs_pinn}(a) shows the solution obtained for the standard PINN (red line), the PINN with DBCs (blue dashed line), and the RK method with exact initial conditions (solid black line). In this example, we consider the interval $x \in [0, 20]$ with initial conditions $y_1(0) = y_2(0) = y_3(0) = 1$. The parameters $a$, $b$, and $c$ are strictly positive and set to $a = 10$, $b = 28$, and $c = \frac{8}{3}$, a classic configuration known to produce chaotic behavior in the Lorenz system. Since this solution demands greater precision, the number of training points is significantly increased to $N = 6000$, compared to the previous examples.

Notably, the standard PINN fails to produce a meaningful solution without the use of DBCs, as its prediction remains stuck at the initial condition for all $x$. This highlights the significant improvement in accuracy achieved by incorporating DBCs into the PINN framework.

To facilitate a clearer comparison, Fig.~\ref{fig:runge_kutta_vs_pinn}(b)–(d) shows the Lorenz trajectory decomposed into its components. These plots demonstrate the good agreement between the PINN with DBCs and the RK solution, with deviations appearing only near the end of the domain. This discrepancy can be partly attributed to the chaotic nature of the Lorenz system, where small inaccuracies in the PINN predictions, such as at the initial point $x_0 = 0$, can grow exponentially over time, eventually leading to significant divergence from the reference trajectory. This becomes evident when the RK method is initialized with the PINN-predicted value at $x_0 = 0$ (orange dots) instead of the prescribed initial conditions. In this case, the agreement between the two solutions improves markedly. This indicates that the PINN accurately captures the structure of the Lorenz solution but corresponds to a slightly shifted initial condition.

\section{Conclusions}
We introduced the DBC method as an improvement to the solutions obtained by PINNs in solving differential equations. In this method, we add additional information gathered from previous training steps to the loss in order to enhance the solution of the PINNs. This approach addresses two key issues in PINNs: their tendency to converge to trivial solutions and to provide misleading results (incorrect predictions despite low losses), especially at points far from the initial conditions. These problems are particularly pronounced in differential equations with rapidly changing terms or strongly oscillatory solutions. Our tests on the optical ray equation and the chaotic Lorenz attractor showed that by incorporating DBCs significantly improves PINN performance, providing a practical and effective strategy for implementing PINNs in complex scenarios.

\hrulefill
\\
\textbf{Funding} The Spanish "Ministerio de Ciencia, Innovaci\'on y Universidade supports and funds this works", MICIU/AEI/10.13039/501100011033 under Grants No. PID2023-148359NB-C21 (also funded by FEDER, UE), No. CEX2023-001286-S (Severo Ochoa Centers of Excellence) and CEX2021-001144-S-20-10 (predoctoral research fellowship). We also acknowledge the Arag\'on Regional Government through project QMAD (\text{E09\_23R}).

\textbf{Disclosures} The authors declare no conflicts of interest.

\textbf{Data availability}
\phantomsection
\label{codigo}
Data underlying the results presented in this paper are publicly available at the following GitHub link :\\\url{https://github.com/AndresMartnz/optical ray equation-in-black-hole-PINN.git.}

\bibliography{sample}

\end{document}